\documentclass[10pt, a4paper, conference, compsocconf]{IEEEtran}
\usepackage[utf8]{inputenc}
\usepackage[T1]{fontenc}
\usepackage{cite}
\usepackage[cmex10]{amsmath}

\usepackage{amssymb}
\usepackage{algpseudocode}
\makeatletter
\let\OldStatex\Statex
\renewcommand{\Statex}[1][3]{%
  \setlength\@tempdima{\algorithmicindent}%
  \OldStatex\hskip\dimexpr#1\@tempdima\relax}
\renewcommand{\ALG@beginalgorithmic}{\small}
\makeatother

\usepackage{url}
\usepackage{colonequals}
\usepackage{graphicx}
\usepackage{pgf}
\usepackage{tikz}
\usepackage{tikz-qtree}
\usetikzlibrary{shapes,arrows,automata,patterns,matrix}
\usetikzlibrary{positioning,trees}
\usetikzlibrary{calc,decorations.markings}
\ifCLASSOPTIONcompsoc
\usepackage[caption=false,font=normalsize,labelfont=sf,textfont=sf]{subfig}
\else
\usepackage[caption=false,font=footnotesize]{subfig}
\fi

\usepackage{listings}
\newsavebox{\mylistingbox}

\newtheorem{definition}{Definition}
\newtheorem{theorem}{Theorem}
\IEEEoverridecommandlockouts

\begin{document}

\title{A Grammatical Inference Approach to Language-Based Anomaly Detection\\in XML}

\author{\IEEEauthorblockN{Harald Lampesberger}
 \thanks{This is a preprint version of the paper accepted for publication at the \textit{First Int. Workshop on Emerging Cyberthreats and Countermeasures}, 
September 2-6, 2013, Regensburg, Germany.}
\IEEEauthorblockA{
Christian Doppler Laboratory for Client-Centric Cloud Computing,\\
Softwarepark 21, 4232 Hagenberg, Austria\\
Email: h.lampesberger@cdcc.faw.jku.at}
}



\maketitle

\begin{abstract}
False-positives are a problem in anomaly-based intrusion detection systems. To counter this issue, we discuss anomaly detection for the eXtensible Markup Language (XML) in a language-theoretic view. We argue that many XML-based attacks target the syntactic level, i.e. the tree structure or element content, and syntax validation of XML documents reduces the attack surface. XML offers so-called schemas for validation, but in real world, schemas are often unavailable, ignored or too general. In this work-in-progress paper we describe a grammatical inference approach to learn an automaton from example XML documents for detecting documents with anomalous syntax.

We discuss properties and expressiveness of XML to understand limits of learnability. Our contributions are an XML Schema compatible lexical datatype system to abstract content in XML and an algorithm to learn visibly pushdown automata (VPA) directly from a set of examples. The proposed algorithm does not require the tree representation of XML, so it can process large documents or streams. The resulting deterministic VPA then allows stream validation of documents to recognize deviations in the underlying tree structure or datatypes.
\end{abstract}

\begin{IEEEkeywords}
intrusion detection; anomaly detection; XML; grammatical inference
\end{IEEEkeywords}

%


\section{Introduction}

Detecting attacks against software is the research field of intrusion detection systems (IDS).
We distinguish IDS techniques into \textit{misuse-} and \textit{anomaly-based} detection for hosts and networks:
A misuse-based IDS matches \textit{signatures} in a stream of events or network traffic.
Contrary, an anomaly-based IDS isolates events or network packets that deviate from \textit{normal behavior}.
While signatures represent known patterns of misbehavior, normality for anomaly detection is usually approximated from observations by machine learning or stochastic methods.

In theory, anomaly detection has the advantage of recognizing yet unknown (\textit{zero-day}) or \textit{targeted attacks} that are specifically designed to evade signatures.
Bilge and Dumitras \cite{Bilge2012} show that zero-day attacks are actually frequent and targeted attacks like Stuxnet \cite{stuxnet} will likely reoccur in the future.
Anomaly detection seems like a perfect solution but suffers from severe practical problems.

False-positives and the high costs associated with them are one major problem \cite{Axelsson1999}.
We don't know beforehand how often attacks occur, so the ratio of normal to abnormal events can be heavily skewed:
Even a system with low false-positive rate could generate an unacceptable number of false-positives.
Sommer and Paxson \cite{Sommer2010} identify further issues why anomaly detection is not adopted outside academia:
It is hard to understand semantics of a detected anomaly and the notion of normality is unstable, especially in networks.
Commercial antivirus and network IDS software still relies on signature-based techniques and anomaly detection only plays a supporting role in products that offer behavioral analysis.

The goal of this paper is a more promising anomaly detection technique for the \textit{eXtensible Markup Language} (XML).
XML is a platform-independent language for semi-structured data and a pillar of today's Web.
Reducing the attack surface therefore makes sense.
For our approach we resort to \textit{formal language theory} and \textit{grammatical inference} to understand language-theoretic properties and learnability of XML.
We believe, a detection technique can only guarantee low false-positive and high detection rates if it respects these properties.

\subsection{Problem Definition}

We consider anomaly detection similar to grammatical inference:
Learning a \textit{representation} of a language, e.g. a grammar or automaton, from the \textit{presentation} of a language, e.g. from examples, counter-examples or an oracle.
Grammatical inference assumes that there is some hidden \textit{target} representation to be discovered, where language class and type of presentation influence successfulness of learning \cite[pp. 141--172]{DelaHiguera2010}.
A learning algorithm is said to converge if the hidden representation is uncovered. 

We define our problem as follows: Given a set of example \textit{XML documents}, a learner returns an automaton that allows \textit{validation} of \textit{syntactic structure} and \textit{datatypes} to decide normality of future documents.

While XML is exchanged as document, the underlying logical model is a \textit{tree}.
Processing the tree as \textit{Document Object Model} (DOM) \cite{w3cdom} requires all the information in memory and this becomes harder with increasing size.
We require both automaton and learner to operate in a \textit{streaming fashion}, where memory and time for processing is limited.
The \textit{Simple API for XML} (SAX) \cite{saxproject} is our streaming interface to documents.

We approach the problem by first discussing expressiveness of XML.
For that, we introduce a formal abstraction of practical schema languages and show that language representation through \textit{visibly pushdown automata} (VPA) is equivalently expressive.
VPA are an executable model capable of stream processing documents and satisfy our requirement.
We then characterize an XML language class that can be efficiently learned from a set of example documents.
Content in XML is from an unknown language class in general.
We therefore introduce a datatype system for abstracting content of possibly infinite nature into a finite set of datatypes.
The contributions are an XML Schema compatible lexical datatype system and a state-merging algorithm for learning VPA.
An inferred VPA can validate future documents to recognize anomalous syntax.

The paper is structured as follows: In the remaining introduction we discuss vulnerabilities, XML-based attacks and introduce our learning setting.
In Section \ref{sec:xml} we define notations, schema languages and analyze XML expressiveness and its limits for stream validation.
VPA are introduced in Section \ref{sec:vpa}.
Section \ref{sec:inference} presents our datatype system and learning algorithm.
Related work is listed in Section \ref{sec:related_work} and Section \ref{sec:conclusion} concludes this paper.

\subsection{A Language-Theoretic View on Security}

Sassaman et al. \cite{Sassaman2011a} analyze the software vulnerability problem using formal language theory.
Modularization and composition is an important process in software engineering but implicitly requires \textit{interfaces} and \textit{protocols} between components.
A protocol basically specifies the syntax and semantics of a formal language for encoding information, e.g. a file format or network message.

When two components $R$ and $S$ interact, the sender $S$ encodes information w.r.t. the protocol as transportable object, e.g. a network message or file.
The receiver $R$ decodes (\textit{parses}) this object according to the protocol and $R$'s internal state is updated in the process.
Unfortunately, protocols in the real world are often ambiguous, under-specified or implementations have errors \cite{Sassaman2011a}.
Sender $S$ might be able to craft a special object such that $R$ moves into an unexpected or insecure state upon parsing.
This object is then called \textit{exploit} because it bends or breaks the original intention of the protocol; We say $S$ abuses a \textit{vulnerability} in the protocol to attack $R$.


An unambiguous and precise protocol specification is required to resolve vulnerabilities such that the receiving component can reject malformed entities \cite{Sassaman2011a}.
This is exactly the \textit{membership decision problem} in formal languages and it may be intractable or undecidable depending on the language class.
Another difficulty is that protocols are often \textit{layered} such that several languages are embedded within each other, e.g. TCP/IP or content in an XML document.

Today's IDS are typically engineered around a specific language class, where computational complexity is tractable.
Nevertheless, their goal is to detect exploits in a possibly larger language class or across several layers of embedded languages.
False-positives and false-negatives are a direct consequence of mismatching language classes.
For example, misuse-based IDS are often restricted to the class of \textit{regular word languages} ($\mathcal{REG}$).
If the class of the observed protocol is greater than $\mathcal{REG}$ and there is a vulnerability, there might be infinite variants of exploits that \textit{evade} signatures over $\mathcal{REG}$.
Understanding the language-theoretic problems is therefore important.

\subsection{Why Secure XML Processing Matters}

XML takes the role of the protocol in Web browsers, mobile applications and Web services.
The logical tree structure allows high expressiveness but correct processing becomes more complex and vulnerabilities arise.
DOM parsers are vulnerable to Denial-of-Service (DoS) attacks that exhaust time and memory, for example by \textit{overlong element names} or \textit{oversized payload}.
A \textit{coercive parsing} attack causes DoS by nesting a vast amount of tags \cite{wsattacks}.

If an XML parser respects the \textit{Document Type Definition} (DTD) in the preamble of a document, several DoS attacks based on \textit{entity expansion} become a threat.
Furthermore, the XML parser could expose confidential information if \textit{external entity references} enable local file import \cite{wsattacks}.

\begin{figure}
	\centering
\begin{lrbox}{\mylistingbox}%
\begin{minipage}{0.49\linewidth}%
\begin{lstlisting}
<transaction>
  <total>1000.00<total>
  <cc>
      1234
  </cc>
</transaction>
\end{lstlisting}

\end{minipage}
\end{lrbox}
	\subfloat[Expected format, attacker controls credit card number \cite{wsattacks}.]{\usebox{\mylistingbox}}
\begin{lrbox}{\mylistingbox}%
\begin{minipage}{0.45\linewidth}%
\begin{lstlisting}
<transaction>
  <total>1000.00<total>
  <cc>
      1234' or '1'='1
  </cc>
</transaction>
\end{lstlisting}
\end{minipage}
\end{lrbox}
	\subfloat[SQL-injection attack.]{\usebox{\mylistingbox}}	
	
\begin{lrbox}{\mylistingbox}%
\begin{minipage}{0.8\linewidth}%
\begin{lstlisting}
<transaction>
  <total>1000.00<total>
  <cc>
      1234</cc><total>1.00</total><cc>1234
  </cc>
</transaction>
\end{lstlisting}
\end{minipage}
\end{lrbox}
	\subfloat[XML injection attack for DOM parsers \cite{wsattacks}.]{\usebox{\mylistingbox}\label{fig:xml_attack_dom}}\hfill

\caption{XML-based attacks.}
\label{fig:att_injection}
\end{figure}

\textit{XML injection} is a large class of XML attacks, where the attacker controls parts of a document.
Figure \ref{fig:att_injection} describes a fictional transaction document, where a monetary amount is given and the user provides a credit card number.
In Figure \ref{fig:xml_attack_dom}, the attacker manipulates the transaction value in the DOM tree when a DOM parser is in place \cite{wsattacks}.
XML injection affects SAX parsers too if the parser state is not propagated correctly.
Cross-Site Scripting in the Web is also a form of injection, where a script or Iframe is embedded.
Classic attacks like \textit{SQL-}, \textit{command-} or \textit{XPATH-injection} are also a threat if the application that utilizes the XML parser is vulnerable.

Note that all the presented example attacks change the \textit{expected syntax} of a document.
Unexpected tree structure or wrong datatypes could lead to harmful interpretation in the XML processing component.
Falkenberg et al. \cite{wsattacks} and Jensen et al. \cite{Jensen2009} recommend \textit{strict validation} of XML documents to mitigate attacks but validation requires a language representation, i.e. a \textit{schema}.

Unfortunately, validation is not common. Only $8.9\%$ of XML documents in the Web refer validate to a schema \cite{Grijzenhout2011}.
Also, Web paradigms like \textit{Asynchronous JavaScript and XML} (AJAX) \cite{ajax} do not enforce schemas or validation, so developers are misled to ad-hoc design.
This motivates learning a language representation from effectively communicated XML for later validation.

\subsection{Learning in the Limit}

We consider Gold's \textit{learning in the limit from positive examples} \cite{Gold1967} as our grammatical inference setting.
The target class, a language class $\mathcal{L}$ expressible by a class of language describing devices $\mathcal{A}$, is \textit{identifiable} in the limit if there exists a learner $I$ with the following properties:
Learner $I$ receives as input enumerated examples $E(1), E(2), \dots$ of some language $L \in \mathcal{L}$, where $E \colon \mathbb{N} \to L$ is an enumeration of $L$, and examples may be in arbitrary order with possible repetitions.
With every input, $I$ returns the current hypothesis $A_i \in \mathcal{A}$, e.g. a grammar or automaton, and there is a point of convergence $N(E)$: For all $j \geq N(E)$, $A_j = A_{N(E)}$ and the language of $A_{N(E)}$ is $L$.
We call $I$ a learner for target class $\mathcal{L}$ if there is convergence for all $L \in \mathcal{L}$.
A sample set $S_+ \subseteq L$ is called \textit{characteristic} if learning converges when $S_+$ is enumerated to $I$ \cite{Fernau2009}.

Unfortunately, grammatical inference is hard and even the class $\mathcal{REG}$ is not learnable in the limit from positive examples only \cite{Gold1967}.
Learning from XML documents is even harder because it is a \textit{context-free word language}. 
Ignorance of learnability properties reflects in bad practical performance of anomaly-based IDS.
We therefore approach the problem more formally and present a learner for a restricted class of XML in Section \ref{sec:inference}.

\section{XML}
\label{sec:xml}

The logical structure of XML is a tree, where $\Sigma$ always denotes the alphabet of element names.
We encode attributes as elements with a leading $@$-character and namespaces as part of the element name.
An encoding example is in Figure \ref{fig:example_xml}.
We disregard identifiers and references because they change the logical structure.

The structure without element content or attribute values is characterized by $\Sigma$-trees \cite{Neven2002a}. The inductive definition of $\mathcal{T}_{\Sigma}$, the set of all $\Sigma$-trees, is:
(1) every $c \in \Sigma$ is a $\Sigma$-tree;
(2) if $c \in \Sigma$ and $t_1, \dots, t_n \in \mathcal{T}_{\Sigma}, n \ge 1$ then $c(t_1, \dots, t_n)$ is a $\Sigma$-tree.
$\Sigma$-trees are \textit{unranked} such that every node can have an arbitrary number of children.

The set of nodes of tree $t \in \mathcal{T}_{\Sigma}$ is $Dom(t) \subseteq \mathbb{N}^*$ and defined as follows: If $t = c(t_1 \cdots t_n)$ with $c \in \Sigma$, $n \geq 0$ and $t_1,\dots,t_n \in \mathcal{T}_{\Sigma}$, then $Dom(t) = \{\epsilon\} \cup \{i.u \mid i \in \{1, \dots, n\}, u \in Dom(t_i)\}$.
Symbol $\epsilon$, the empty word, is the root of the tree and node $v.j$ is the $j$-th child of node $v$. The label of $v$ in $t$ is $lab^{t}(v)$.
A \textit{tree language} $L$ over $\Sigma$ is then a set of trees such that $L \subseteq \mathcal{T}_{\Sigma}$.

A document is a $\Sigma$-tree encoded with tags.
For notational convenience, we strip angled brackets such that the set of open-tags is $\Sigma$ and
the set of close-tags becomes $\overline{\Sigma} = \{ \overline{c} \mid c \in \Sigma\}$.
We use variables $c, c_1, c_2, \dots , c_i \in \Sigma$ for open-tags and $\overline{c}, \overline{c_1}, \overline{c_2}, \dots , \overline{c_i} \in \overline{\Sigma}$ for the according close-tags.
XML documents without content are words over $(\Sigma \cup \overline{\Sigma})$ and \textit{well-matched} if they obey the grammar $W \coloncolonequals WW \mid c W \overline{c} \mid \epsilon$.

A tree is translated into a document by pre-order traversal (document order) and we denote the function $doc$ as the bijection between trees and documents.

\subsection{Schemas and Types}

A schema is a tree grammar that restricts expressible XML over elements $\Sigma$ and implicitly gives meaning to structure.
DTDs are the simplest form of schemas:
\begin{definition}[DTD \cite{Martens2006}]
A DTD is a triple $(\Sigma, d, s_d)$, where production rules $d \colon \Sigma \to \mathcal{REG}(\Sigma)$ map element names to \textit{regular expressions} over $\Sigma$ and $s_d$ is the distinguished start element.
The right-hand side of production rules is called \textit{content model} and $L(d)$ is the set of trees that satisfy $d$.
\end{definition}

The expressible language class $\mathcal{DTD}$ is rather limited and practical schema languages like XML Schema (XSD) \cite{w3cschema} or Relax NG \cite{relaxng} offer \textit{types} to increase expressiveness.
Types are from a finite set, each type is associated with a unique element name and the start element has exactly one type \cite{Martens2006}.
Variables $m,m_0,n$ in this paper always denote types.
As a formal abstraction of practical schema languages we recall the definition of \textit{extended DTD} (EDTD):

\begin{definition}[EDTD \cite{Martens2006}]
An EDTD $D$ is a tuple $D = (\Sigma, M, d, m_0, \mu)$, where $M$ is a set of types, $\mu \colon M \to \Sigma$ is a surjection from types onto element names and $(M, d, m_0)$ is a DTD over types.
A tree $t$ satisfies $D$ if $t=\mu(t')$ for some $t' \in L(d)$, where $\mu$ ranges over trees.
Tree $t'$ is called \textit{witness} for $t$, $L(D)$ denotes the set of trees and $L^w(D)$ denotes the documents that satisfy $D$.
The language class $\mathcal{EDTD}$ expressible by EDTDs is equivalent to the \textit{regular tree languages} \cite{Neven2002a}.
\end{definition}


\subsection{Stream Validation and Expressiveness}

The stream validation or type-checking problem is to decide whether a document is in the language of a given schema within a single pass.
Typing a document $w$ is to assign every position $i$ (every tag) some type.
A document is \textit{valid} w.r.t. a schema if such an assignment is possible for all positions.
Note that function $\mu$ is surjective and a position can have multiple types in general.

Martens et al. \cite{Martens2006} and Murata et al. \cite{Murata2005} discuss \textit{ambiguity} and \textit{determinism} of schemas: A schema is ambiguous if there is a document in the language with multiple types at some position.
A schema is \textit{deterministic} if for all described documents at all positions the choice is limited to a single type.
In other words, a schema is deterministic if every type assignment is clear when an open-tag is read.
Note that ambiguity always implies nondeterminism.
Martens et al. \cite{Martens2006} introduce the \textit{1-pass pre-order typed} (1PPT) property for EDTDs that are deterministic and therefore allow efficient stream validation.

Note that determinism is also important for efficient processing.
This is one factor why content models in practical schema languages like DTD and XSD have restrictions to enforce determinism \cite{w3cschema}.
We direct the reader to Martens et al. \cite{Martens2006} for a thorough analysis of expressiveness of schemas.

We have $\mathcal{DTD} \subsetneq \mathcal{EDTD}^{st} \subsetneq \mathcal{EDTD}^{rc} \subsetneq \mathcal{EDTD}$, where $\mathcal{EDTD}^{st}$ is the class of schemas that satisfy the restrictions of XSD and $\mathcal{EDTD}^{rc}$ is the class of deterministic schemas, where the 1PPT property holds \cite{Martens2006}.

\subsection{Datatypes and Mixed Content}

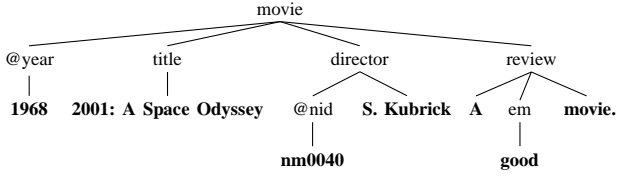
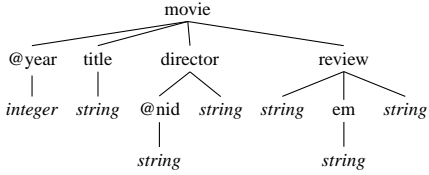
\begin{figure}
	\centering
\begin{lrbox}{\mylistingbox}%
\begin{minipage}{0.99\linewidth}%
\begin{lstlisting}
<movie year="1968">
  <title>2001: A Space Odyssey</title>
  <director nid="nm0040">S. Kubrick</director>
  <review>A <em>good</em> movie.</review>
</movie>
\end{lstlisting}
\end{minipage}
\end{lrbox}

	\begin{tikzpicture}
		\tikzstyle{every node}=[font=\scriptsize]
		\tikzset{grow=down,level distance=1.9em}
		\node (f1) at (0,2.9) {\subfloat[XML document with attributes, data and mixed content.]{\label{fig:example_xml_doc}\usebox{\mylistingbox}}};
		\node (f2) at (0,0) {\subfloat[Tree with data nodes.]{
		\label{fig:example_xml_tree}
		\Tree [.movie
		        [.@year \textbf{1968} ]
				[.title \textbf{2001: A Space Odyssey} ]
				[.director [.@nid \textbf{nm0040} ] \textbf{S. Kubrick} ]
				[.review \textbf{A } [.em \textbf{good} ] \textbf{ movie.} ]
				]}};
		\node (f3) at (0,-3.3) {\subfloat[Datatyped tree.]{
		\label{fig:example_xml_dttree}
		\Tree [.movie
		        [.@year \textit{integer} ]
				[.title \textit{string} ]
				[.director [.@nid \textit{string} ] \textit{string} ]
				[.review \textit{string} [.em \textit{string} ] \textit{string} ]
				]}};
	\end{tikzpicture}
\caption{Example XML document and it's tree representation.}
\label{fig:example_xml}
\end{figure}

XML documents carry data as element contents or attribute values and the tag encoding of documents guarantees that tags and data are not confused.
We denote data in a document as words over alphabet $U$, i.e. Unicode, and variables $r, s \in U^*$ always represent data.
But data could be from any language class, for example natural language or program code.
As an abstraction of data, we introduce \textit{datatypes}:

\begin{definition}
A so-called lexical \textit{datatype system} is a tuple $(\Delta, U, \phi)$, where $\Delta$ is a finite set of datatypes and $\phi \colon \Delta \to \mathcal{P}(U^*)$ is a surjection that assigns every datatype its \textit{lexical space} as some language over $U$.
Because it is surjective, a datum $r$ may have several matching datatypes.
Variables $a,b \in \Delta$ always denote datatypes in this paper.

A datatype system has functions $types \colon U^* \to \mathcal{P}(\Delta)$ and $firstType \colon U^* \to \Delta$.
While $types(r)$ returns all matching datatypes for some $r$, $firstType(r)$ chooses one matching datatype to reduce arbitrary data to datatypes.
\end{definition}

With respect to the tree structure, content $r$ between tags $cr\overline{c}$ is encoded into a single child node $v.j$ representing the datatype, where $lab^t(v) = c$ and $lab^t(v.j) = r$.
We call such a node \textit{data node} for short.
Data nodes are always tree leafs and we now refine EDTDs with datatypes:

\begin{definition}
A \textit{datatype extended DTD} ($\Delta$-EDTD) is a tuple
$D = (\Sigma, M, d, m_0, \mu, \Delta, U, \phi)$, where elements $(\Sigma, M, d, m_0, \mu)$ form an EDTD and $(\Delta, U, \phi)$ is a datatype system.
As an extension to EDTDs, production rules $d \colon M \to \mathcal{REG}(M \cup \Delta)$ assign every type a content model as regular expression over both types and datatypes.
Production rules only allow expressions, where a datatype is followed by either a type or $\epsilon$, but never a subsequent datatype.
Lexical spaces $\phi \colon \Delta \to \mathcal{REG}(U)$ are restricted to regular expressions over $U$.

A \textit{datatyped tree} $t'$ over $(\Sigma \cup \Delta)$ satisfies $D$ if $t' = \mu(t'')$ for some $t'' \in L(d)$, where $\mu$ applies only to elements.
We denote $L_\Delta(D)$ as the set of datatyped trees that satisfy $D$.
A tree with data nodes $t$ satisfies $D$ if there is a datatyped tree $t' \in L_\Delta(D)$ such that $lab^t(u) \in \phi(lab^{t'}(u))$ holds for all data nodes $u$ and $lab^t(v) = lab^{t'}(v)$ holds for all nodes $v$ with $lab^{t'}(v) \in \Sigma$.
Language $L(D)$ denotes the set of trees with data nodes that satisfy $D$.
\end{definition}

Accordingly, we define the word languages generated by $\Delta$-EDTD $D$.
Suppose that bijection $doc$ transforms trees with data nodes and datatyped trees into documents like in Figure \ref{fig:example_xml}.
Then $L_\Delta^w(D) = \{doc(t) \mid t \in L_\Delta(D) \}$ is the \textit{datatyped word language} and the \textit{document (word) language} generated by $D$ is $L^w(D) = \{doc(t) \mid t \in L(D) \}$.

$\Delta$-EDTDs allow so-called \textit{mixed content}, where between two tags both data and other nested tags are allowed.
Mixed content typically appears in markup languages, e.g. the XML Hypertext Markup Language (XHTML).
Regarding structural expressiveness, the language classes of EDTDs also translate to our definition of $\Delta$-EDTDs.
We now have an abstraction of schema languages that captures attributes, datatypes and mixed-content on a syntactic level.

%
%
%
%

\section{Visibly Pushdown Automata for XML}
\label{sec:vpa}

The well-matched tags in documents induce a visible nesting relation.
In fact, XML is a \textit{visibly pushdown language} (VPL) \cite{Alur2004} and Kumar et al. \cite{Kumar2007} show that every EDTD-definable document language is a VPL.
This property holds for our definition of $\Delta$-EDTDs because tags are still well-matched and we encode attributes as nested elements.
VPLs are accepted by \textit{visibly pushdown automata} (VPA), a restricted form of pushdown automata, where the input symbol determines the stack action.

\begin{definition}[VPA \cite{Alur2004}]
$A = (\tilde{\Sigma}, Q, q_0, Q^F, \Gamma, \delta)$ is a VPA, where $\tilde{\Sigma} = (\Sigma_{call}, \Sigma_{int}, \Sigma_{ret})$ is the pushdown alphabet made of three distinct alphabets, $Q$ is the set of states, $q_0 \in Q$ is the start state, $Q_F \subseteq Q$ are the final states, $\Gamma$ is the stack alphabet and the transition relation is $\delta = \delta^{call} \cup \delta^{int} \cup \delta^{ret}$, where $\delta^{call} \subseteq (Q \times \Sigma_{call} \times Q \times \Gamma)$, $\delta^{int} \subseteq (Q \times \Sigma_{int} \times Q)$ and $\delta^{ret} \subseteq (Q \times \Gamma \times \Sigma_{ret} \times Q)$.
\end{definition}

A transition $(q, c, q', \gamma) \in \delta_{call}$, denoted as $q \xrightarrow{c/\gamma} q'$, is a call-transition from state $q$ to $q'$ that pushes $\gamma$ on the stack when symbol $c \in \Sigma_{call}$ is read. A transition $(q, \gamma, \overline{c}, q') \in \delta_{ret}$, written as $q \xrightarrow{\overline{c}/\gamma} q'$, is a return-transition from state $q$ to $q'$ that pops $\gamma$ from the stack when symbol $\overline{c} \in \Sigma_{ret}$ is read. An internal transition $(q, a, q') \in \delta_{int}$, denoted as $q \xrightarrow{a} q'$, moves from state $q$ to $q'$ at input $a \in \Sigma_{int}$ without changing the stack.
We direct the reader to Alur and Madhusudan \cite{Alur2004} for the semantics of VPA.

Contrary to traditional pushdown automata, VPA can be determinized and are closed under complement, intersection, union, concatenation and Kleene-star. Also language equivalence, emptiness, universality and inclusion are decidable.

Next we will show the equivalence of $\Delta$-EDTDs and XML VPA (XVPA) \cite{Kumar2007}.
XVPA are a special form of \textit{modular VPA} that go back to program modeling.
In a modular VPA, states are partitioned into \textit{modules} and the stack alphabet is exactly the set of states.
When a module calls another one, the current state is saved on the stack and popped for returning.
With respect to XML, modules are exactly the types.
Call, return and internal transitions of the VPA are the open-tag, close-tag and character events of the SAX interface to documents.

We assume the following about SAX: There exists a global datatype system $(\Delta, U, \phi)$ and every datum $r$ between two tags or attribute value is reduced to one character event.
For stream validation, the SAX interface reports only the first matching datatype $firstType(r)$ to the XVPA instead of $r$ for efficiency.
So, the XVPA processes \textit{datatyped documents} and the internal alphabet over datatypes is guaranteed to be finite.

\begin{definition}[XVPA \cite{Kumar2007}]
An XVPA $A$ is a tuple $A \!\!=\!\! (\Sigma, \Delta, M, \mu, \{(Q_m, e_m, X_m, \delta_m)\}_{m \in M}, m_0, F)$, where $\Sigma$, $\Delta$, $M$ and $\mu$ have the same meaning as in $\Delta$-EDTDs, $m_0$ is the distinguished start type and $F = X_{m_0}$ are final exit states. Every type $m \in M$ characterizes a module, where
\begin{itemize}
	\item $Q_m$ is the finite set of module states,
	\item $e_m \in Q_m$ is a single entry state of the module,
	\item $X_m \subseteq Q_m \mbox{ is the exit of module } m$ (exit states),
	\item Transitions $\delta_m = \delta_m^{call} \cup \delta_m^{ret} \cup \delta_m^{int}$, where
	\begin{itemize}
		\item $\delta_m^{call} \subseteq \{q_m \xrightarrow{c/q_m} e_n \mid n \in \mu^{-1}(c)\}$,
	    \item $\delta_m^{ret} \subseteq \{q_m \xrightarrow{\overline{c}/p_n} q_n \mid q_m \in X_m \land n \in \mu^{-1}(c) \}$ and is deterministic, i.e. $q_n = q_n'$ whenever $q_m \xrightarrow{\overline{c}/p_n} q_n$ and $q_m \xrightarrow{\overline{c}/p_n} q_n'$, and
   		\item $\delta_m^{int} \subseteq \{q \xrightarrow{a} q' \mid q,q' \in Q_m \land a \in \Delta\}$.
	\end{itemize}
\end{itemize}
\end{definition}

Return transitions are always deterministic by definition. The XVPA is \textit{deterministic} if also the call transitions are deterministic.
The semantics of an XVPA are given by its corresponding VPA $A' = (\tilde{\Sigma}, Q, q_0, \{q_f\}, Q, \delta)$, where $\tilde{\Sigma} = (\Sigma, \Delta, \overline{\Sigma})$, $q_0$ and $q_f$ are start and accepting state, $Q = \{q_0, q_f\} \cup \bigcup_{m \in M} Q_m$ and transition function $\delta$ is defined as
\[
\delta \! = \!\!\! \bigcup_{m \in M} \!\! \delta_m \cup \{q_0 \xrightarrow{\mu(m_0)/q_0} \!e_{m_0}\} \cup \{q \xrightarrow{\overline{\mu(m_0)}/q_0} \!q_f |\; q \!\in\! F\}.
\]

The language $L_A(m)$ of module $m$ is a \textit{datatyped word language} and accepted words are of form $\mu(m) w \overline{\mu(m)}$.
The accepted language $L(A) = L_A(m_0)$ of XVPA $A$ is the datatyped word language $L(A')$ of its corresponding VPA.

The set $X_m$ are exit states, where at least one return transition originates from.
In a valid XVPA, the single-exit property \cite{Kumar2007} must hold:
If there is some return transition $q_m \xrightarrow{\overline{c}/p_n} q_n$ from module $m$ to $n$, then there must be return transitions $q_m' \xrightarrow{\overline{c}/p_n} q_n$ for all exit states $q_m' \in X_m$.
The single-exit property guarantees that $L_A(m)$ is always the same, independent from the calling state or module.

\begin{theorem}\label{thm:rdtdxvpa}
Given a datatype system $(\Delta, U, \phi)$, every $\Delta$-EDTD $D$ has a corresponding XVPA $A$ such that their datatyped word languages are equal $L(A) = L_{\Delta}^w(D)$. Also, for every XVPA $A$ there is an equivalent $\Delta$-EDTD $D$ such that $L_\Delta^w(D) = L(A)$.
\end{theorem}

The proof is skipped, it refines Kumar et al. \cite{Kumar2007} with datatypes.
Intuitively, every type $m$ has an intermediate DFA $D_m$ in the translation between XVPA modules and regular expressions $d(m)$ in $\Delta$-EDTD production rules.
The time complexity of the automaton for processing a document is linear in the length of the document and the required space is bounded by the nesting depth.

\section{Inference from Streaming XML}
\label{sec:inference}

The goal is to learn an XVPA from a set of example documents $S_+$.
Inference boils down to (1) defining a datatype system, (2) characterizing types and states and (3) learning the language over types and datatypes from examples in $S_+$.

\subsection{XML Schema Compatible Lexical Datatype System}

\begin{figure}
\centering
\begin{tikzpicture}[->,>=stealth',shorten >=1pt,auto,semithick]
\tikzstyle{every node}=[font=\footnotesize\itshape]
\node (v1) at (-4.7,2.7) {booleanNum};
\node (v3) at (-9.5,3.5) {boolean};
\node (v2) at (-4.7,3.5) {unsignedByte};
\node (v4) at (-3.5,3.5) {byte};
\node (v5) at (-9.5,7.3) {language};
\node (v6) at (-6,8) {NCName};
\node (v7) at (-8.3,7.3) {duration};
\node (v9) at (-8.3,3.9) {dateTimeDuration};
\node (v8) at (-7,4.6) {yearMonthDuration};
\draw  (v1) edge (v2);
\draw  (v1) edge (v3);
\draw  (v1) edge (v4);
\draw  (v3) edge (v5);
\draw  (v5) edge (v6);
\draw  (v7) edge (v6);
\draw  (v8) edge (v7);
\draw  (v9) edge (v7);
\node (v10) at (-6,8.8) {QName};
\node (v11) at (-6,9.6) {Name};
\node (v12) at (-6,10.4) {NMTOKEN};
\node (v13) at (-6,11.2) {token};
\node (v14) at (-8.5,11.5) {normalizedString};
\node (v15) at (-6,11.9) {string};
\draw  (v6) edge (v10);
\draw  (v10) edge (v11);
\draw  (v11) edge (v12);
\draw  (v12) edge (v13);
\draw  (v13) edge (v14);
\draw  (v14) edge (v15);
\node (v17) at (-3.5,11.6) {base64BinaryLF};
\node (v16) at (-3.5,10.8) {base64Binary};
\draw  (v16) edge (v17);
\draw  (v17) edge (v15);
\draw  (v16) edge (v13);
\node (v18) at (-3.6,9.6) {gMonth};
\node (v19) at (-2.4,9.6) {gDay};
\node (v20) at (-8.8,10.1) {gMonthDay};
\node (v21) at (-2.6,10.1) {gYearMonth};
\draw  (v18) edge (v12);
\draw  (v19) edge (v12);
\draw  (v20) edge (v12);
\draw  (v21) edge (v12);
\node (v24) at (-4.8,9.6) {double};
\node (v23) at (-4.8,8.8) {decimal};
\node (v22) at (-4.8,8) {integer};
\draw  (v22) edge (v23);
\draw  (v23) edge (v24);
\draw  (v24) edge (v12);
\node (v25) at (-4.7,4.4) {unsignedShort};
\node (v26) at (-4.7,5.2) {unsignedInt};
\node (v27) at (-4.7,6) {unsignedLong};
\node (v28) at (-5.7,6.8) {nonNegativeInteger};
\draw  (v2) edge (v25);
\draw  (v25) edge (v26);
\draw  (v26) edge (v27);
\draw  (v27) edge (v28);
\draw  (v28) edge (v22);
\node (v29) at (-3.5,4.4) {short};
\node (v30) at (-3.5,5.2) {int};
\node (v31) at (-3.5,6) {long};
\draw  (v4) edge (v29);
\draw  (v29) edge (v30);
\draw  (v30) edge (v31);
\draw  (v31) edge (v22);
\node (v34) at (-2.4,7.3) {gYear};
\node (v33) at (-2.7,6.8) {nonPositiveInteger};
\node (v32) at (-2.5,2.3) {negativeInteger};
\draw  (v32) edge (v33);
\draw  (v33) edge (v22);
\draw  (v34) edge (v22);
\node (v35) at (-6.7,6) {evenLenInteger};
\draw  (v35) edge (v28);
\node (v36) at (-7.1,7.3) {hexBinary};
\draw  (v35) edge (v36);
\draw  (v36) edge (v6);
\node (v41) at (-7,9.6) {anyURI};
\node (v38) at (-8.8,9.6) {dateTime};
\node (v39) at (-9,8.4) {dateTimeStamp};
\node (v40) at (-7.6,8.4) {time};
\node (v37) at (-6.8,8.4) {date};
\draw  (v37) edge (v38);
\draw  (v39) edge (v38);
\draw  (v40) edge (v38);
\draw  (v41) edge (v12);
\draw  (v10) edge (v41);
\draw  (v38) edge (v12);
\draw  (v2) edge (v29);
\draw  (v25) edge (v30);
\draw  (v26) edge (v31);
\node (v44) at (-3.5,2) {boolean0};
\node (v42) at (-5.7,2) {boolean1};
\node (v43) at (-5.7,3.9) {positiveInteger};
\draw  (v42) edge (v43);
\draw  (v43) edge (v28);
\draw  (v44) edge (v33);
\draw  (v44) edge (v1);
\draw  (v42) edge (v1);
\end{tikzpicture}
\caption{Poset of lexical datatypes $\Delta$ in lexical inclusion order.}
\label{fig:datatypesystem}
\end{figure}
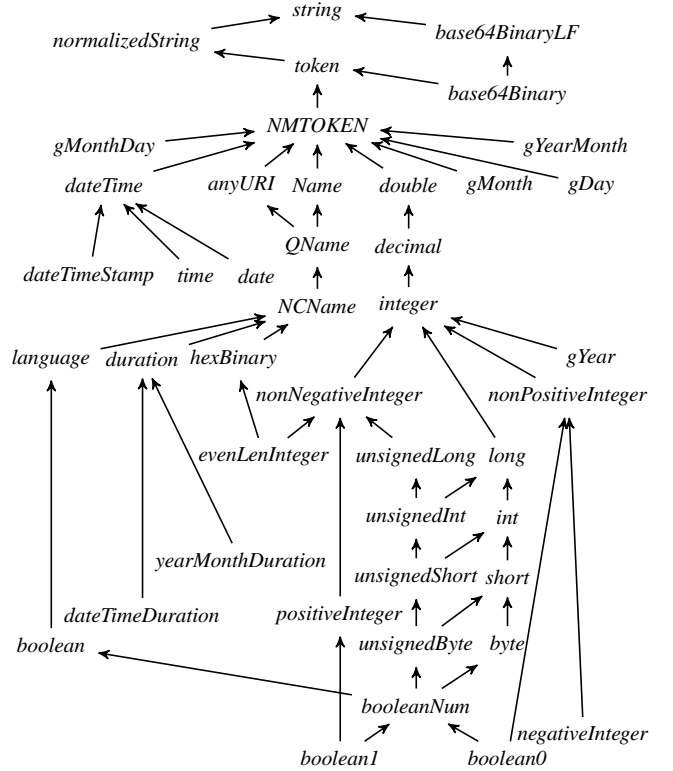

XVPA in our definition use datatypes as finite alphabet for internal transitions.
In Section \ref{sec:xml} we introduce the notion of datatype system but for inference we need a concrete instance.
XSD defines a rich set of 47 atomic datatypes together with a hierarchy \cite{w3cdatatypes}, where every datatype has a semantical value space and a lexical space that is characterized by a regular expression.
Unfortunately, the lexical spaces of XSD datatypes heavily overlap, for example the word '0' is in the lexical space of datatypes \textit{boolean}, \textit{Integer}, or \textit{string} to name a few.
A learner only experiences the lexical space and this leads to the problem of choosing the correct datatype for a set of words.

Let $(\Delta, U, \phi)$ be our datatype system, where $U$ is the Unicode alphabet.
Based on XSD datatypes we define $\Delta$ as a \textit{poset} of 44 datatypes and it is shown in Figure \ref{fig:datatypesystem}.
The partial order is the subset relation $\subseteq$ over individual lexical spaces, i.e.
\[
a, b \in \Delta \colon a \leq b \iff \phi(a) \subseteq \phi(b) \mbox{ , where}
\]
surjection $\phi$ maps datatypes to the lexical space definitions of XSD \cite{w3cdatatypes} respectively.
The following adoptions to datatypes and lexical spaces are made:
\begin{itemize}
	\item Datatype \textit{anyURI} has an unrestricted lexical space in the XSD standard.
	In our definition, a datum has datatype \textit{anyURI} iff it is a RFC 2396 \textit{Unified Resource Identifier} with a defined scheme and path.
	\item The exponents of datatype \textit{double} are unrestricted.
	\item Datatypes \textit{float, IDREF, IDREFS, ENTITY, ENTITIES, ID, NOTATION} and \textit{NMTOKENS} are dropped because their lexical spaces are indistinguishable from others.
	\item We add \textit{boolean0, boolean1, booleanNum, evenLenInteger} and \textit{base64BinaryLF} to resolve some severe ambiguities.
\end{itemize}

If some content $r$ matches datatype $a$ then it also matches datatypes $b_1, b_2, \dots, b_n$ iff $a \leq b_i$ for $1 \leq i \leq n$.
So, the best characterization of $r$ is its \textit{minimal datatype}.
We refine functions $types$ and $firstType$ to reflect the partial order: Function $types(r)$ returns all matching minimal datatypes and $firstType(r)$ chooses one matching minimal datatype.
For inference, we define the inverse closure $cl^{-1} \colon \mathcal{P}(\Delta) \to \mathcal{P}(\Delta)$ that returns all datatypes that are smaller or equivalent than a given set of datatypes.

\subsection{Characterizing Types and States}

Martens et al. \cite{Martens2006} characterize types in XSD based on ancestors.
Given document $w$ and position $i$, then the ancestor string $anc\mbox{-}str(w, i) = c_1 c_2 \cdots c_j$ is the string of unmatched open-tags in the document prefix $w_{1,i} = c_1w_1c_2w_2 \dots c_jw_j$.
A schema has \textit{ancestor-based types} if there exists a hypothetical function $f \colon \Sigma^* \to M$ that assigns every open-tag at position $i$ in document $w$ a single type $f(anc\mbox{-}str(w_{1,i}, i))$.
This restriction is exactly the Element Declarations Consistent (EDC) rule of XSD \cite{w3cschema}.
Identifying types is then defining relation $\sim_M$ that partitions $\Sigma^*$ into equivalence classes of ancestor strings.
Note that we restrict our learning algorithm automatically to a subset of language class $\Delta\mbox{-}\mathcal{EDTD}^{st}$ by assuming that types are ancestor-based.

Regarding content models we know that the full class of regular languages is not learnable from positive examples \cite{Gold1967}.
Bex et al. \cite{Bex2004} show that the majority of regular expressions in real world schemas are in fact simple such that every type occurs at most $k$ times in an expression ($k$-ORE).
The language of a $k$-ORE is a $(k+1)$-\textit{testable regular language}, where grammatical inference from positive examples is feasible \cite{Garcia1990}.

Our learning strategy is \textit{state-merging}: We first construct a specific VPA that represents exactly $S_+$ and then generalize by merging similar states.
We denote pairs $(x, y) \subseteq (\Sigma^* \times (\Sigma \cup \{\$\})^*)$ as VPA states, where $x$ is an ancestor string and $y$ is a left sibling string $lsib\mbox{-}str(w, i)$.
Symbol $\$ \notin \Sigma$ denotes a placeholder for XML content and $lsib\mbox{-}str(w, i) = d_1c_1d_2c_2 \cdots d_{n-1}c_{n-1}d_{n}$, where $c$ is the rightmost unmatched open-tag in the document prefix $w_{1,i} = ucd_1c_1v_1\overline{c_1}d_2c_2v_2\overline{c_2} \cdots d_{n-1}v_{n-1}c_{n-1}\overline{c_{n-1}}d_{n}$, $d_1, d_2, \dots, d_n \in \{\$,\epsilon\}$ are optional placeholders and the well-matched substrings $c_jv_j\overline{c_j}$ for $1 \leq j < n$ represent sibling nodes in the tree w.r.t. to position $i$. As an example, suppose $w$ is the document in Figure \ref{fig:example_xml_doc} and position $i$ is just before tag $\langle/$\textit{review}$\rangle$ then $lsib\mbox{-}str(w, i) = \$ \cdot em \cdot \$$.

\subsection{The Learning Algorithm}

\begin{figure}
\begin{algorithmic}[1]
\Function{DtVPPA$^{st}$}{$S_+$}
	\State \textbf{global} $(\Delta, U, \phi)$
	\Comment{datatype system}
    \State $\Sigma \gets Q_F \gets \delta^{call} \gets \delta^{ret} \gets \delta^{int}  \gets \emptyset$ 		\Comment{initialization}
    \State $t \colon Q \times Q \to \mathcal{P}(\Delta)$
    \Comment{empty dictionary}
    \State $q_0 \gets (\epsilon, \epsilon)$
    \State $Q \gets \{q_0\}$
    \ForAll{$w \in S_+$}
    \Comment{iterate over documents}
    \State $stack \gets [\bot]$
    \State $q \gets q_0$
    \ForAll{$(event, data) \in SAXEvents(w)$}
	    \If {$startElement(event)$}
	    \Comment{open-tag}
	    	\State $\Sigma \gets \Sigma \cup \{data\}$
	    	\State $push(stack, q)$
	    	\State $q' \gets (\pi_1(q) \cdot data, \epsilon)$
	    	\State $\delta^{call} \gets \delta^{call} \cup \{q \xrightarrow{data/q} q'\}$
	    \ElsIf{$endElement(event)$}
	    \Comment{close-tag}	    
	    	\State $assert(data = \pi_{-1}(\pi_1(q)))$
	    	\Comment{matching?}
	    	\State $p \gets pop(stack)$
	    	\State $q' \gets (\pi_1(p), \pi_2(p) \cdot data)$
	    	\State $\delta^{ret} \gets \delta^{ret} \cup \{q \xrightarrow{\overline{data}/p} q'\}$
	    \ElsIf{$characters(event)$}
  	    \Comment{content}
	    	\State $q' \gets (\pi_1(q), \pi_2(q) \cdot \$)$
         	\State $t(q, q') \gets t(q, q') \cup types(data)$
	    \EndIf
	    \State $Q \leftarrow Q \cup \{q'\}$
	    \State $q \leftarrow q'$
    \EndFor
    \State $Q_F \leftarrow Q_F \cup \{q\}$
    \EndFor
    \State $\delta^{int} = \{q \xrightarrow{a} q' \mid a \in cl^{-1}(t(q, q'))\}$
    \Comment{int. transitions}
    \State \Return $((\Sigma, \Delta, \overline{\Sigma}), Q, q_0, Q_F, Q, \delta^{call} \cup \delta^{int} \cup \delta^{ret})$
\EndFunction
\end{algorithmic}
\caption{Visibly Pushdown Prefix Acceptor for class $\Delta\mbox{-}\mathcal{EDTD}^{st}$.}
\label{alg:vppa-st}
\end{figure}

Intuitively, the inference algorithm (1) constructs a so-called \textit{visibly pushdown prefix acceptor} (VPPA) from the sample set, (2) merges similar states, (3) partitions states into modules, (4) adds missing return transitions to satisfy the single-exit property of XVPA and (5) minimizes the XVPA by merging equivalent modules.
Figure \ref{alg:inferxvpa-st} gives the full algorithm.

With $\pi_1,\pi_2, \dots, \pi_n$ we denote projections of the first, second and $n$-th element and $\pi_{-1}$ is the last element of a tuple or word 
A VPPA is a deterministic VPA that represents exactly the examples from $S_+$ and construction requires only a single pass.
The idea of a VPPA is that every prefix of every document in $S_+$ leads to a unique state in the automaton, similar to a prefix tree acceptor \cite[p. 238]{DelaHiguera2010}.
Algorithm $\textsc{DtVppa}^{st}$ is listed in Figure \ref{alg:vppa-st}.
While iterating over documents in $S_+$, the algorithm remembers all datatypes that occur between two states in a dictionary-like data structure.
After iteration, internal transitions are added for all datatypes in the inverse closure of remembered datatypes.
This guarantees that during stream validation the automaton allows a transition if the first matching minimal datatype returned by SAX is valid.

\begin{figure}
\begin{algorithmic}[1]
\Function{InferDtXVPA$^{st}_{k,l}$}{$k, l, S_+$}
	\State \textbf{global} $(\Delta, U, \phi)$
	\Comment{datatype system}
    \State $((\Sigma, \_, \_), Q, q_0, Q_F, Q^F, \delta) \gets \textsc{DtVppa}^{st}(S_+)$
    \While{$\exists q_1,q_2 \in Q \colon q_1 \sim_{k,l} q_2$}
    \Comment{state merging}
    	\State $mergeStates(f_{k,l}, q_1, q_2)$
    \EndWhile
    \State $M \gets \{\pi_1(q) \mid \mbox{for all } q \in Q \land \pi_1(q) \neq \epsilon\}$
    \State $m_0 \gets \pi_1(\delta(q_0, c, q_0))$
    \Comment{module called by $q_0$}
    \ForAll{$m \in M$}
    \Comment{XVPA conversion}
    	\State $e_m \gets (m, \epsilon)$
    	\State $Q_m \gets \{q \in Q \mid \pi_1(q) = m\}$
    	\State $\delta_m \gets \{rel \in \delta \mid \pi_1(rel) \in Q_m\}$
    	\State $X_m \gets \{q_m \mid \exists c,p_n, q_n \colon (q_m \xrightarrow{\overline{c}/p_n} q_n) \in \delta_m \}$
    	\State $\delta_m \leftarrow \delta_m \cup \{q_m \xrightarrow{\overline{c}/p_n} q_n \mid  \mbox{for all } q_m \in X_m$
    	\Statex[5.6] $\mbox{ if } \exists q_m' \colon (q_m' \xrightarrow{\overline{c}/p_n} q_n) \in \delta_m \}$
    \EndFor
    \While{$\exists m,n \in M \colon m \sim_M n$}
    \Comment{minimization}
    	\State $mergeModules(m, n)$
    \EndWhile
    \State $\mu \gets \{ m \mapsto c \mid m \in M \land \exists q \colon (q \xrightarrow{c/q} e_m) \in \delta^{call} \}$
    \State \Return $(\Sigma, \Delta, M, \mu, \{(Q_m, e_m, X_m, \delta_m)\}_{m \in M},$
    \Statex[3.3] $m_0, X_{m_0})$
\EndFunction
\end{algorithmic}
\caption{The learning algorithm returns an XVPA with datatypes.}
\label{alg:inferxvpa-st}
\end{figure}

Merging states in the second step generalizes the VPPA.
Function $f_{k,l} \colon (\Sigma^* \times (\Sigma \cup \{\$\})^*) \to (\Sigma^{\leq l} \times (\Sigma \cup \{\$\})^{\leq k})$ is a so-called \textit{distinguishing function} \cite{Fernau2003} that restricts a state $q$ to its local neighborhood by stripping down $\pi_1(q)$ to its $l$-length suffix and $\pi_2(q)$ to its $k$-length suffix.
With respect to $f_{k,l}$, two states are similar $q_1 \sim_{k,l} q_2$ if they map to the same state $f_{k,l}(q_1) = f_{k,l}(q_2)$.
The single state $f_{k,l}(q_i)$ represents equivalence class $[q_i]_{\sim_{k,l}}$, all states in the equivalence class and their transitions are merged into the representative and the VPA stays deterministic.

In the third step, the VPA is turned into an XVPA by partitioning all states $q \in Q$ based on their ancestor-string component $\pi_1(q)$.
Types then are $M \subseteq \Sigma^{\leq l}$ and algorithm $\textsc{DtVppa}^{st}$ guarantees that $(m, \epsilon)$ is the single entry state of every module $m$.
Start type $m_0$ is the one called from state $(\epsilon, \epsilon)$ and the module of type $\epsilon$ is ignored.
The XVPA does not satisfy the single-exit property yet.
Let $X_m$ be all module states, where some return transition originates from.
We add missing returns such that every module $n$ calling $m$ experiences the same language $L_A(m)$.

In the last step, the XVPA is minimized by merging equivalent modules.
We define equivalence relation $\sim_M$ such that types $m$ and $n$ are the same if their modules are called by the same open-tag and their corresponding DFA $D_m$ and $D_n$ as constructed in the proof of Theorem \ref{thm:rdtdxvpa} are equivalent.
If $m \sim_M n$ we redirect all calls and returns from $n$ to $m$ and remove $n$.
Finally, $\mu$ maps all types to the elements they are called by.
Note that learning the VPPA and state merging can be combined into one efficient step.

\begin{figure*}
\centering
\begin{tikzpicture}[->,>=stealth',shorten >=1pt,auto,semithick]
\tikzstyle{every node}=[font=\small]
\tikzstyle{sm} = [circle, draw, inner sep=2pt, label distance=-4pt]
\tikzstyle{sm2} = [circle, draw, inner sep=0pt,minimum size=12pt]
\node[draw=none] (dummy1) at (-4.1,1) {};
\node[sm,label={$\epsilon, \epsilon$}] (ee) at (-3.5,1) {};
\node[sm,label=180:{$a, \epsilon$}] (ae) at (-3.5,0) {};
\node[sm,label=180:{$aa, \epsilon$}] (aae) at (-3.5,-1) {};
\node[sm,label=180:{$aa, \$$}] (aa2) at (-3.5,-2) {};

\node[sm,label=90:{$a, a$}] (aa) at (-2.3,0) {};
\node[sm,label=0:{$ab, \epsilon$}] (abe) at (-1.3,-1) {};
\node[sm,label=180:{$ab, \$$}] (abd) at (-1.3,-2) {};

\node[sm,label=0:{$a, abb$}] (aabb) at (-1.3,0) {};
\node[sm,label=0:{$a, ab$}] (aab) at (-0.3,-2) {};
\node[sm,double, label={$\epsilon, a$}] (ea) at (-1.3,1) {};
\path (dummy1) edge (ee)
(ee) edge node[left] {$a$} (ae)
(ae) edge node[left] {$a$} (aae)
(aae) edge node[left] {$decimal_\Delta$} (aa2)
(aa2) edge node {$\overline{a}$} (aa)
(aa) edge node [pos=0.2] {$b$} (abe)
(abe) edge node[left] {$string_\Delta$} (abd)

(abd) edge node [below] {$\overline{b}$} (aab)
(aab) edge node[above, pos=0.3] {$b$} (abe)
(abe) edge node [right] {$\overline{b}$} (aabb)
(aabb) edge node {$\overline{a}$} (ea);

\node (dummy2) at (1.2,1) {};
\node[sm,label={$\epsilon, \epsilon$}] (ree) at (1.8,1) {};
\node[sm,label=180:{$a, \epsilon$}] (rae) at (1.8,0) {};
\node[sm,label=180:{$aa, \epsilon$}] (raae) at (1.8,-1) {};
\node[sm,label=270:{$aa, \$$}] (raa2) at (1.8,-2) {};
\node[sm,label={$a, a$}] (raa) at (3,0) {};
\node[sm,label=180:{$ab, \epsilon$}] (rabe) at (3.7,-1) {};
\node[sm,label=0:{$ab, \$$}] (rabd) at (4.5,-2) {};

\node[sm,label=0:{$a, b$}] (rab) at (4.5,0) {};
\node[sm,double,label={$\epsilon, a$}] (rea) at (4.5,1) {};

\path (dummy2) edge (ree)
(ree) edge node {$a$} (rae)
(rae) edge node[left] {$a$} (raae)
(raae) edge node[left] {$decimal_\Delta$} (raa2)
(raa2) edge node {$\overline{a}$} (raa)
(raa) edge node [left]{$b$} (rabe)
(rabe) edge node[left] {$string_\Delta$} (rabd)

(rabd) edge node[right] {$\overline{b}$} (rab)
(rab) edge[bend left=10, pos=0.7] node {$b$} (rabe)
(rabe) edge[bend left=10] node {$\overline{b}$} (rab)
(rab) edge node {$\overline{a}$} (rea);

\node[font=\normalsize] at (3.2,1.7) {\textbf{Merged States}};
\node[font=\normalsize] at (-2.3,1.7) {\textbf{VPPA}};
\node[font=\normalsize] at (8.3,1.7) {\textbf{XVPA}};

\node[sm2] (xae) at (8.4,0.4) {$\epsilon$};
\node[sm2] (xaa) at (8.4,-0.6) {$a$};
\node[sm2,double] (xab) at (8.4,-1.6) {$b$};
\node[sm2] (xaae) at (6.5,0.4) {$\epsilon$};
\node[sm2,double] (xaa2) at (6.5,-0.6) {$\$$};
\node[sm2,double] (xabe) at (10.6,-0.6) {$\epsilon$};

\node (dummy3) at (11.4,0.8) {};
\node[sm2] (q0) at (10.6,0.8) {$q_0$};
\node[sm2,double] (qf) at (6.5,-1.6) {$q_f$};

\node[sm2,double] (xabg) at (10.6,-1.6) {$\$$};
\draw [very thick] (7.9,1) rectangle (8.9,-2);
\draw  (5.5,1) rectangle (7,-1);
\draw  (10.1,0.1) rectangle (11.6,-2);
\node at (8.4,0.8) {$\mathbf{a}$};
\node at (6.2,0.8) {$\mathbf{aa}$};
\node at (10.9,-0.1) {$\mathbf{ab}$};

\path (xae) edge node [above] {$a/\epsilon_a$} (xaae)
(xaae) edge node [left] {$dec._\Delta$} (xaa2)

(xaa2) edge node {$\overline{a}/\epsilon_a$} (xaa)
(xaa) edge [bend left=50, pos=0.5] node  {$b/a_a$}(xabe)
(xabe) edge node  {$str._\Delta$} (xabg)
(dummy3) edge (q0)
(q0) edge[dashed] node [above]  {$a/q_0$} (xae)
(xab) edge[dashed] node [above] {$\overline{a}/q_0$} (qf)

(xabg) edge [bend left=50, pos=0.5]  node {$\overline{b}/a_a$} (xab)
(xabe) edge [bend left=10, pos=0.75] node {$\overline{b}/b_a$} (xab)
(xab) edge [bend left=20, pos=0.75] node {$b/b_a$}(xabe);

\end{tikzpicture}
\caption{$\textsc{InferDtXVPA}^{st}_{k,l}$ example for $S_+ = \{aa\mathbf{10.0}\overline{a}b\mathbf{TEXT}\overline{b}b\overline{b}\overline{a}\}$ and parameters $k=1, l=2.$}
\label{fig:inferexample}
\end{figure*}

\subsection{Example and Discussion}
Figure \ref{fig:inferexample} gives a toy example, where the sample set holds a single document.
The SAX interface abstracts the contents $\mathbf{10.0}$ and $\mathbf{TEXT}$ into simplified datatypes $decimal_\Delta$ and $string_\Delta$ respectively.
Note that the state $(ab, \epsilon)$ is visited twice during the VPPA construction because both open-tags $b$ in context of element $a$ have the same ancestor-string $ab$. 
During state merging, the states $(a, ab)$ and $(a, abb)$ are collapsed into the single state $(a,b)$. 
In the example, the parameter $l=2$ leads to two different types $\mathbf{a}$ and $\mathbf{aa}$ in the final XVPA.
While both corresponding modules are called by the same tag $a$, they have completely different content models.

The parameters $k$ and $l$ constrain locality of a state.
The language class $\Delta\mbox{-}\mathcal{EDTD}^{st}_{k,l} \subsetneq \Delta\mbox{-}\mathcal{EDTD}^{st}$ is learnable if $k$ and $l$ are bound and $S_+$ is \textit{characteristic} such that every valid transition in the XVPA appears at least once in the set.
Unfortunately, we do not know whether a sample set is characteristic.
But we can guarantee that the quality of the learned automaton stays the same or improves with every example in the sample set if the hidden target is in language class $\Delta\mbox{-}\mathcal{EDTD}^{st}_{k,l}$.

If $l=1$ then types are exactly element names and the algorithm learns a proper subset of $\Delta\mbox{-}\mathcal{DTD}$.
A parameter $k = 1$ limits the left-sibling string of a state to element names or the $\$$ symbol, so inferred XVPA modules become equivalent to Single Occurrence Automata \cite{Bex2010} in terms of expressiveness.
In the case that $k$ and $l$ are chosen too small, the resulting automaton over-generalizes the language.
Contrary, increasing the parameters requires much larger characteristic sets for convergence.

\section{Related Work}
\label{sec:related_work}
XML stream validation is first discussed by Segoufin and Vianu \cite{Segoufin2002}. Kumar et al. \cite{Kumar2007} introduce VPA as executable model for XML that captures the entire class of regular tree languages. Schewe et al. \cite{Schewe2009} extend VPA for approximate XML validation and Picalausa et al. \cite{Picalausa2011} present an XML Schema framework using VPA.

For a survey of grammatical inference we direct the reader to the book of de la Higuera \cite{DelaHiguera2010}.
Fernau \cite{Fernau2003} introduces function distinguishable languages and we apply this concept in Section \ref{sec:inference} for state merging.
Kumar et al. \cite{Kumar2006} mention that query learning VPA with counterexamples is possible but our setting is different.

Several results on DTD inference from XML have been published \cite{Bex2010,Bex2010a,Fernau2001,Garofalakis2003}, but we aim for the strictly larger class of XSDs.
Ml\'{y}nkov\'{a} \cite{Mlynkova2008} presents a survey of XSD inference.
The general idea is to start with an extended context-free grammar as schema abstraction, inferred from examples, and merge non-terminals \cite{Mlynkova2009}.
Hegewald et al. \cite{Hegewald2006} and Chidlovskii \cite{Chidlovskii2001} also handle datatypes in their presented methods.
Our approach is similar to Bex et al. \cite{Bex2007}. Their algorithms use tree automata for learning $l$-local Single Occurrence XSDs in a probabilistic setting but without datatypes.

In the field of information retrieval, Kosala et al. \cite{Kosala2006} and Raeymaekers et al. \cite{Raeymaekers2008a} give algorithms to infer HTML wrappers as tree automata.
Regarding intrusion detection, Rieck et al. \cite{Rieck2009} introduce approximate tree kernels as a similarity measure for trees and use them for anomaly detection in HTML.

To our knowledge the presented approach is the first that directly learns an automaton model with both streaming and datatypes in mind.
A hard problem in learning schemas is to find nice regular expressions for content models.
We focus on learning an automaton representation and intentionally leave conversion to regular expressions open, as many of the noted references propose heuristics or solutions.

\section{Conclusion and Future Work}
\label{sec:conclusion}

We approached the problem of anomaly detection in XML more formally and introduced $\Delta$-EDTDs as abstraction of practical schema languages with datatypes. We showed that XVPA are an equivalent model capable of stream validation and contributed a lexical datatype system and an algorithm for learning an XVPA from a set of documents.
The algorithm converges for target class $\Delta\mbox{-}\mathcal{EDTD}^{st}_{k,l}$ given the sample set is characteristic.
A learned automaton could theoretically be converted into an XSD schema.

The presented work is still in an early stage.
We already have a working prototype which is our baseline for further research and the next step is a thorough evaluation with XML-based attacks.
First experiments with the prototype indicate that abstraction by the lexical datatype system using XSD datatypes is too coarse in some cases.
We will therefore look into approximations of specific datatypes during learning.
Other improvements are to extend the learnable language class and redefine the algorithms for incremental learning.
Also, we do not know if some sample set is characteristic and leads to convergence. A refinement to a probabilistic learning setting could enhance applicability when sample sets are incomplete or noisy.

Finally, it is of great interest how our approach to XML inference and stream validation translates to other prominent semi-structured languages like JSON or HTML.
An application in mind is a client-side component that learns how Web applications and services communicate with a Web client and detects syntactical deviations, for example caused by Cross-Site Scripting attacks.

\section*{Acknowledgment}
We thank Philipp Winter for the helpful feedback and suggestions.
This research has been supported by the Christian Doppler Society.

\bibliographystyle{IEEEtran}
\bibliography{IEEEabrv,pub_bibliography,urls}

\end{document}